# Measurement of mass by optical forced oscillation of absorbing particles trapped in air

Jinda Lin[1*], Jianliao Deng[1], Rong Wei[1], Yong-qing Li[2], and Yuzhu Wang[1]

[1] Key Laboratory of Quantum Optics and Center for Cold Atom Physics, Shanghai Institute of Optics and Fine Mechanics, Chinese Academy of Sciences, 201800, China

[2]Department of Physics, East Carolina University, Greenville, North Carolina 27858-4353, USA

*Corresponding author.   jinda@siom.ac.cn

We demonstrate the measurement of mass of the absorbing micro-particle trapped in air by optical forced oscillation. When the trapping light intensity is modulated sinusoidally, the particle in the trap undergoes forced oscillation and the amplitude of the oscillation depends directly on the modulated frequency. Based on a simple spring model, we fit the amplitudes versus the modulated frequencies and obtain the stiffness of the optical trap and the mass of the trapped particle. The fitting results show that, for a certain particle, the stiffness varies linearly with the trapping light intensity while the mass is consistent. The density of the micro-particle is then estimated and could be used to classify different kinds of absorbing particles, like C and CuO.

## 1.INTRODUCTION

Optical trapping or optical tweezers technology has become an important tool for manipulating micro- and nano-scale particles since the pioneering experiments of Ashkin [1]. Combined with the Raman spectroscopy, optical tweezers becomes a powerful analytic tool, termed as Raman tweezers, for the rapid identification of biological, chemical, or pharmaceutical particles in solution and in air [2-5]. Coupled with advanced position detection systems, optical tweezers provides precise measurements of pico-Newton force, displacement, and instantaneous velocity of microscopic system [6-8]. To study the interaction between different kinds of samples and measure the mechanical properties of cells, optical forced oscillation has been implemented in optical tweezers [9,10]. Even more, the fundamental properties of the micro- and nano-sized particles trapped by optical tweezers, like the radius and the temperature, can be measured by analyzing the power spectral density (PSD), the velocity autocorrelation function, or the mean-square displacement of the Brownian motion [11,12].

Optical trapping of airborne absorbing micro-particles by photophoresis has been successfully demonstrated using simple experimental configurations [3,13,14]. Since the photophoretic forces are orders of magnitude larger than radiation pressure forces for an absorbing particle, it can levitate and stably trap nigrosin, Johnson grass smut spores, riboflavin, and carbon black in air without the use of tight focusing objective lenses, which cannot be done by using radiation pressure force or gradient force [15]. Photophoresis is also important for various fields of aerosol science, like levitation in the stratosphere and planet formation [16,17]. Therefore, lots of great efforts are made to clarify the origination, calculation, determinant, and application of photophoresis, as has been reviewed by ref 16. However, compared to the extensive study of trapping using optical gradient forces in solution, attentions are merely paid on the behavior and properties of the micro-particles trapped in air by photophoretic force. Recently, the rotation of the trapped absorbing particle was measured directly by a high-speed imaging technique using a charge coupled device (CCD) camera and pulsed illumination or indirectly by a single photodiode [18]. Base on the light intensity scattered from the rotated particle, the Brownian motion of the trapped particle was measured and the mass of the particle was estimated by analyzing the PSD of the Brownian motion [19].

In this letter, we demonstrate the measurement of mass of the absorbing particle trapped in air by a loosely focused Gaussian beam. We modulated the trap light intensities with sinusoidal function and the particles in the trap undergo optical forced oscillations, the amplitudes of which depend directly on the modulated frequency. After the roll-off frequency, the amplitudes of the oscillation decrease quickly when the modulated frequencies increase. By fitting the amplitudes versus the modulated frequencies through the simple spring model, the stiffness and the mass are measured. The fitting results show the stiffness changes linearly with the trap light intensities while the mass is constant for different light intensities. Compared to the measurement of mass by analyzing the Brownian motion in ref. [19], we use a single CCD camera to record the trajectory and avoid detecting the modulated scattered light due to the rotation of the particles. The particle moves back and forth and the air temperature along the trajectory is the room temperature.

## 2. EXPERIMENTAL SETUP

The schematic of the experimental setup is shown in Fig.1(a). A continuous-wave laser beam at 780 nm is focused horizontally by a lens ($f_0$=5 cm) to form the trap. The light beam is linearly polarized in $TEM_{00}$ mode and expanded to about 6 mm in diameter. In the experiment, we use an acousto optical modulator (AOM) in the light path to change the light intensity $I$. The trapping intensity is then modulated by changing the driving voltage of the AOM with the form of

$I=I_0+M\sin(\omega t)$, where $I_0$ is the trapping laser intensity before

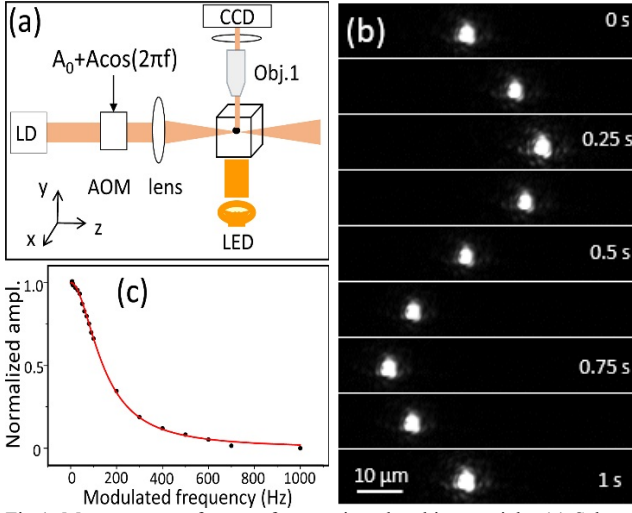

Fig.1. Measurement of mass of a trapping absorbing particle. (a) Schematic setup for single particles trapped by a single Gaussian beam. LD — diode-pumped laser, AOM — acousto optical modulator, Obj. — objective lens, CCD — charge coupled device, LED — light-emitting diode. (b) Sequential images of the scattered light of a modulated trapped particle recorded with a CCD camera. The time interval of the images is 125 ms with the driving frequency f=1 Hz and the modulation depth M=2 mW while the trapping intensity is $I_0$ =20 mW. The scale bar is 10 μm. (c) Normalized amplitude versus modulated frequency under the trapping intensity of 20 mW with modulated depth of 2 mW.

modulation, $M$ is the modulated depth, and $\omega = 2\pi f$ is the circular frequency of the modulation. The light-absorbing particles we used in this experiment are irregular particles of ground HB pencil lead powder or cluster of CuO powder and they are gently blown into the laser beam. The particles are trapped inside a cuvette near the focus of the lens and imaged through a long working distance objective (20×, NA=0.4) onto a CCD camera. The position of the trapped particle can be manipulated by a few millimeters in three dimensions by a translation stage that mounts the lens, so that the particle can be positioned in the focal plane of the objective for clear image. A light-emitting diode is used for illumination. Fig.1 (b) depicts the sequential images of a trapped particle under forced oscillation with modulated frequency of 1 Hz. The images from up to down were recorded with time interval of about 125 ms. The trapping light intensity is 20 mW with the modulated depth of 2 mW. The amplitude of the forced oscillation can then be measured from Fig.1 (b). After calibration, the amplitude of the forced oscillation is measured to be 19.6 μm. We then change the modulated frequencies (between 1 and 1000 Hz) and measure the corresponding amplitudes. The results are shown in Fig.1(c) (Normalized by 19.6μm with black dots). The shortest interval of the CCD camera between two images is 1 ms and it is too slow to record the instantaneous positions of the particles when the particles move fast under high modulated frequency. Therefore, for high modulated frequencies (>200 Hz), we set the exposure time of the CCD to 50 ms and the CCD camera records the trajectory in one image. Following the same procedure, the amplitudes of the forced oscillations at different light intensities and modulated depths are measured.

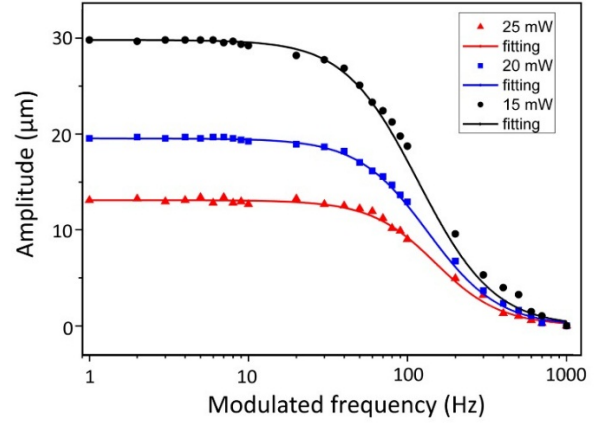

Fig.2 Normalized amplitude versus modulated frequency at the optical trapping power of 15 (black circles), 20 (blue squares), and 25 (red diamonds) mW. The modulation depths are 2 mW for all of the three different light intensities. The solid lines are the theoretical fits according to Eq.(5).

### 3. EXPERIMENTAL RESULTS AND ANALYSIS

Fig. 2 shows the amplitudes of optical forced oscillations versus the modulated frequencies under different trapping laser powers of 15 (black circles), 20 (blue squares), and 25 (red diamonds) mW with same modulated depth of 2 mW. During the oscillation, the photophoretic force provides as the restoring force. The stiffness of the optical trap originating from the photophoretic force is then linearly proportional to the intensity of the trapping laser [19]. Thus, a simple spring model can be used to depict the optical forced oscillation when the trapping intensity is sinusoidally modulated. The equation of motion for the modulated trap is

$$m\ddot{x} + \beta\dot{x} + [k_0 + k_1\sin(\omega t)]x = 0 \qquad (1)$$

where m is the mass of the trapped particle, $x$ is the center position of the particle, $\beta = 6\pi\eta r$, $\eta$ is the viscosity of the surrounding median, and r is the size of the particle, $k_0$ is the stiffness of the trap at the intensity of $I_0$ and $k_1$ arises from the modulation depth $M$. Ignoring the transient solution and the higher harmonic frequency components[9], the steady-state solution for Eq.(1) is

$$A(\omega) = \frac{\sqrt{k_0^2+k_1^2}\, l}{\sqrt{(k_0-m\omega^2)^2+\omega^2\beta^2}} \qquad (2)$$

where $l$ is a constant displacement when the modulating frequency approximate to zero. At low frequency zone, Eq.(2) can be simply expressed as $A = \sqrt{1+(k_1/k_0)^2}\, l$, which means the amplitude decreases when $I_0$ is increased. As shown in Fig.2, the amplitudes of oscillation are 29.8, 19.6, and 13.1 μm for 15 (black), 20 (blue) and 25 (red) mW at modulated frequency of 1 Hz, respectively. For simplicity, we use the normalized amplitude as

$$\frac{A(\omega)}{A(0)} = \frac{k_0}{\sqrt{(k_0 - m\omega^2)^2 + \omega^2\beta^2}} \quad (3)$$

The square of right hand side in Eq.(3) represents a Lorentz profile in the frequency domain. The concept of roll-off frequency $f_c$ [20] is used to determine the stiffness of the trap as

$$k_0 = 2\pi\beta f_c \quad (4)$$

By using Eq.(4), the normalized amplitude becomes

$$\frac{A(f)}{A(0)} = \frac{f_c}{\sqrt{(f_c - \frac{2\pi m}{\beta}f^2)^2 + f^2}} \quad (5)$$

The roll-off frequencies for trapping light intensity of 25, 20, and 15 mW are obtained directly as 76.1, 67.2, and 59.2 Hz from Fig.2, respectively. By fitting the data in Fig.2 with Eq.(5), the parameter $\frac{2\pi m}{\beta}$ is measured. We replace the normalized amplitude A(0) with A(1) for the reason that, the amplitudes equal to the same value when the modulated frequency is less than 10 Hz. Table.1 shows the fitting results of $\frac{2\pi m}{\beta}$ and $fc$ at different trapping laser powers. Noted that the roll-off frequencies can also be inferred directly from Fig.2 or calculated by fitting.

Table.1 Fitted values of the parameters at different trapping powers

| $I_0$(mW) | 15 | 20 | 25 |
|---|---|---|---|
| $fc$ (Hz) | 59.2 | 67.2 | 76.1 |
| $\frac{2\pi m}{\beta}$ | $3.53*10^{-3}$ | $3.59*10^{-3}$ | $3.70*10^{-3}$ |

Up to now, we have not make any assumption on the particle or the around air environment. From Eq.(4) and table 1, the stiffness $k_0$ and the mass $m$ both depend on the viscosity factor β = 6πηr, which is decided by the size of the trapped particle. If the trapped particles are spherical, r equals to the radius. The stiffness and the mass can be calculated with high accuracy. While for the irregular particles we used in the experiments, r is difficult to be exactly measured but directly affects the accuracy of the evaluation of mass and stiffness through β. We take an approximation and assume the shape of the particle is cylindrical. Even though the assumption will bring systematic errors to the calculation of the viscosity and the volume, the demonstration of our method is still valid. Under this assumption, we measure the size of the particle byusing the image with calibration as shown in the inset of Fig.3. to be 4.5μm. The particle is trapped in the air and oscillates back and forth at room temperature of 300 K. The viscosity is then $\eta$=1.86× $10^{-5}$ Pa・s and $\beta = 1.58 \times 10^{-9}$ Pa・s・m. With the fit values shown in Table 1, the stiffness and mass are calculated and shown in Fig.3. Stiffness increases linearly with the laser power as it has been predicted while the mass is constant for the same particle at different trapping power. Totally, the main errors arise from the evaluation of particle size, fitting uncertainties of the oscillation amplitude, and the viscosity.

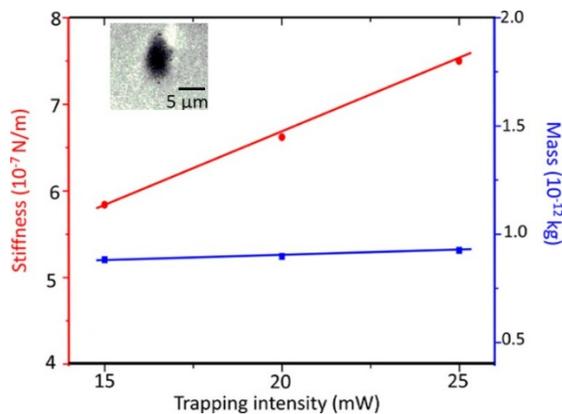

Fig.3 The fitting results of the stiffness (red circles) and the mass (blue squares) for the same particle at different trapping power of 15, 20 and 25 mW. Inset is the CCD image of the trapping particle in the yz-plane.

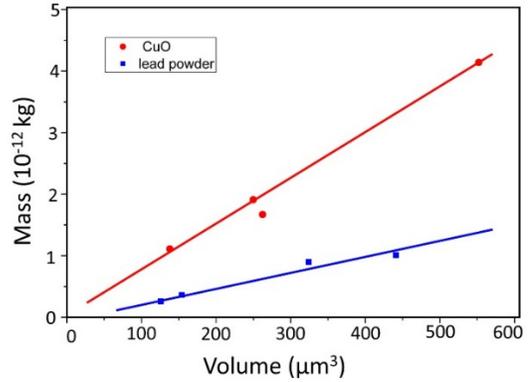

Fig.4 Mass versus volume of different kinds of particles: ground pencil lead powder (blue squares) and CuO (red circles). The slope of blue line is 2.4 g/cm³ for pencil lead powder and the slope of the red line is 7.4 g/cm³ for CuO.

Base on the size and the mass we have calculated using the aforementioned procedure, we can infer the density of the particle and used to classify the particles coarsely. The absorbing particles we used in the experiments are ground HB pencil lead powder or cluster of CuO powder. Shown in Fig.4 are the density of four different particles of pencil lead powder (blue squares) and CuO (red circles). The average densities for pencil lead powder and CuO are 2.4 and 7.4 g/cm³.They are a little different with the real density values for the reason that the particles are irregular and the evaluation of the volumes have systematic errors. However, these errors are indiscriminate for both the pencil lead powder and the CuO, which means the classification is not very unreasonable.

### 4.CONCLUSION
We have demonstrate the measurement of mass of the trapped absorbing particles using optical forced oscillation method. By fitting the amplitude of the oscillation versus the modulated frequency base on the spring model, the stiffness and the mass are measured. The fitting results show the stiffness changes linearly with the trap light intensities while the mass is constant. The density of the trapped particle is then estimated and used to coarsely classify the particles.

**Funding.** This work was supported by the National Nature Sciences Foundation of China under Grant No. 11504393 and No. 91536220; Foundation from Shanghai Science and Technology Committee under Grant No. 15YF1413400; Scientific Research Foundation for the Returned Overseas Chinese Scholars from State Education Ministry under Grant No. 1502411F30. Jinda Lin thanks Xiaolin Li for useful help in preparing the experiment.